\documentclass[epj,twocolumn]{webofc}
\usepackage[varg]{txfonts}   
\usepackage{graphicx}
\usepackage{upgreek}

\wocname{epj}
\woctitle{Seismology of the Sun and the Distant Stars 2016}
\begin{document}
\title{On the relation between activity-related frequency shifts and the sunspot\\ distribution over the solar cycle 23}
%

\author{\firstname{\^{A}ngela R. G.} \lastname{Santos}\inst{1,2,3}\fnsep\thanks{\email{asantos@astro.up.pt}} \and
        \firstname{Margarida S.} \lastname{Cunha}\inst{1,2} \and
        \firstname{Pedro P.} \lastname{Avelino}\inst{1,2} \and
        \firstname{William J.} \lastname{Chaplin}\inst{3,4} \and
        \firstname{Tiago L.} \lastname{Campante}\inst{3,4}
}

\institute{Instituto de Astrof\'{i}sica e Ci\^{e}ncias do Espa\c{c}o, Universidade do Porto, CAUP, Rua das Estrelas, PT4150-762 Porto, Portugal
			  \and Departamento de F\'{i}sica e Astronomia, Faculdade de Ci\^{e}ncias, Universidade do Porto, Rua do Campo Alegre 687, PT4169-007 Porto, Portugal
			  \and School of Physics and Astronomy, University of Birmingham, Edgbaston, Birmingham B15 2TT, UK
			  \and Stellar Astrophysics Centre (SAC), Department of Physics and Astronomy, Aarhus University, Ny Munkegade 120, 8000 Aarhus C, Denmark
		  }

\abstract{The activity-related variations in the solar acoustic frequencies have been known for 30 years. However, the importance of the different contributions is still not well established. With this in mind, we developed an empirical model to estimate the spot-induced frequency shifts, which takes into account the sunspot properties, such as area and latitude. The comparison between the model frequency shifts obtained from the daily sunspot records and those observed suggests that the contribution from a stochastic component to the total frequency shifts is about 30\%. The remaining 70\% is related to a global, long-term variation. We also propose a new observable to investigate the short- and mid-term variations of the frequency shifts, which is insensitive to the long-term variations contained in the data. On the shortest time scales the variations in the frequency shifts are strongly correlated with the variations in the total area covered by sunspots. However, a significant loss of correlation is still found, which cannot be fully explained by ignoring the invisible side of the Sun when accounting for the total sunspot area. We also verify that the times when the frequency shifts and the sunspot areas do not vary in a similar way tend to coincide with the times of the maximum amplitude of the quasi-biennial variations found in the seismic data.}
\maketitle
%
\section{Introduction}
\label{sec:intro}

In the Sun, the activity-related frequency shifts are one of the numerous manifestations of the solar magnetic activity. As the activity level increases, the frequencies of the acoustic modes are observed to increase \cite[e.g.][]{Woodard1985,Libbrecht1990a,Elsworth1990,Chaplin1998}. In addition to this 11-yr, long-term variation, the solar acoustic frequencies also vary on a quasi-biennial timescale \cite[e.g.][]{Fletcher2010,Broomhall2012,Broomhall2015}.

The acoustic frequencies can be affected by different phenomena associated to both weak and strong components of the magnetic field \cite[e.g.][]{Tripathy2007,Jain2009,Broomhall2015,Santos2016}. Those may contribute to the observed frequency shifts through indirect (thermal and structural) effects and  direct (via Lorentz force) effects. Although the activity-related frequency shifts have been known for more than 30 years, the relative impact of the different contributions to the frequency shifts is not fully understood.

In the work presented in this proceedings, we have studied the short-term variations (on a timescale of days) in the solar acoustic frequencies. We may expect that the dominant contribution to such short-term variations is related to sunspots (strong component of the magnetic field), which, depending on their size, usually live and evolve in a few days. In a first study, we estimate the contribution from the spot component to the observed frequency shifts. To that end and following the approach of \cite[][]{Cunha2000}, we derived a model for the spot-induced frequency shifts (Sect.~\ref{sec:sfshifts}, see also \cite{Santos2016}). Sect.~\ref{sec:WD} shows the results from a second study aimed at the characterization of the differences between the observed frequency shifts and the total area covered by sunspots. We have proposed a new observable, the weighted sum of the frequency-shift differences, to study the short- and, also, the mid-term variations in the acoustic frequencies \cite[see][]{Santos2016b}. This quantity is mostly insensitive to the long-term (11-yr) signal.

\section{Model frequency shifts}
\label{sec:sfshifts}

By changing the medium where the acoustic waves propagate, sunspots can affect their propagation properties, in particular the oscillation frequencies. Sunspots contribute to the frequency shifts with both indirect and direct effects. By solving the incomplete wave equations, \cite{Santos2012} studied the spots' indirect effect due structural changes within a sunspot. Here, we consider the total spot impact, by parameterizing the model frequency shifts.

One may quantify the spots' perturbation through the phase difference, $\Delta\delta$, between the magnetic (within a spot) and the non-magnetic wave solutions (in the quiet-Sun), at a depth $R^*$ below the region of influence of the spot (here, we consider a depth of 10 Mm). In this work, we assume a characteristic phase difference, $\Delta\delta_{\rm ch}$, induced by sunspots. The comparison between the model and observed frequency shifts will allow to constrain this parameter.

Following a variational principle \cite{Cunha2000}, taking the area, $A_i$, and central colatitude, $\theta_i$, of a given spot $i$, we find the following relation for the spot-induced frequency shifts over time \cite[for details, see][]{Santos2016}
\begin{equation}
\dfrac{\delta\omega_{lm}}{\omega_{lm}}(t)\simeq-\dfrac{\Delta\delta_{\rm ch}}{\omega_{lm}^2 R^2\int_{r_1^l}^{R^*}c^{-2}\kappa^{-1}dr}\sum_{i=1}^{N(t)}\left[\left(P_l^m\left(\cos\theta_i\right)\right)^2A_i\right],\label{eq:fshifts}
\end{equation}
where $\omega$, $l$, and $m$ are, respectively, the mode frequency, angular degree, and azimuthal order, $R$ is the solar radius, $r_1^l$ is the lower turning point of the mode, $c$ is the sound speed, $\kappa$ is the radial component of the acoustic wave number, $r$ is the distance to the solar center, $N$ is the number of spots at a given time $t$, and $P_l^m$ are the normalized \cite[see][]{Santos2016} Legendre polynomials.

Having the sunspot parameters, Eq. (\ref{eq:fshifts}) can be used to estimate the spot-induced frequency shifts for different radial orders, angular degrees, and azimuthal orders.

Since, our goal is to determine the contribution from the spot component, which varies on a timescale of days, we shall consider frequencies obtained with a relative short cadence. We use the frequency shifts computed by \cite{Tripathy2011} with a cadence of 36 days and an overlap of 18 days, constituting two samples of independent data points. The observed frequency shifts where obtained from GONG (Global Oscillation Network Group) data and includes modes of degree between $l=0$ and $100$ with frequencies ranging from $2000$ and $3300\,\upmu{\rm Hz}$. Within this frequency range, the frequency shifts increase almost linearly with the frequency \cite[e.g.][]{Libbrecht1990a,Chaplin2001}. Thus, we take the central frequency to estimate the mean frequency shifts.
Using the sunspot daily records from the National Geophysical Data Center (NGDC/NOAA) database, the spot-induced frequency shifts, $\delta\nu_{lm}=\delta\omega_{lm}/(2\pi)$, are computed for $l=0-100$ and corresponding azimuthal orders. The mean spot-induced frequency shifts, $\delta\nu_{\rm spots}$, are then obtained by combining $\delta\nu_{lm}$ as the observed frequency shifts \cite[see][]{Tripathy2011,Santos2016}.
Finally, in order to obtain model frequency shifts consistent with the observational data, we average the model spot-induced frequency shifts, obtained from the daily sunspot records, over bins of 36 days with an overlap of 18 days.

To account for the weak component of the magnetic field we consider a global smooth component varying on the 11-yr timescale, given by
\begin{equation}
\delta\nu_{\rm global}=wf(t),
\end{equation}
where $w$ is the weight of the global component and
\begin{equation}
f(t)=\dfrac{a\left(t-t_0\right)^3}{\exp\left(\left(t-t_0\right)^2/b^2\right)-c}.
\end{equation}
This functional form was proposed by \cite{Hathaway1994} to describe the variation of the sunspot number over the cycle. The parameters $a,\,b,\,c$, and $t_0$ are related to the amplitude, starting time, size of the rising phase, and temporal asymmetry of the cycle, respectively, and are given by the best fit to the observed frequency shifts.

The total model frequency shifts are then
\begin{equation}
\delta\nu_{\rm model}=\delta\nu_{\rm global}+\delta\nu_{\rm spots}.\label{eq:mfshifts}
\end{equation}
This way, the model frequency shifts contain two parameters, $w$ and $\Delta\delta_{\rm ch}$, to be determined.

The best model is obtained through a $\chi^2$ minimization between the observed and model frequency shifts. The value found for the characteristic phase difference induced by spots is $\Delta\delta_{\rm ch}\sim-0.44$. The weight of the global component, corresponding to the best fit, is $w\sim0.71$, meaning that the global long-term component accounts for $\sim70\%$ of the total frequency shifts. The remaining $\sim30\%$ are related to the spot-induced component which varies on short timescales.

\begin{figure*}[t]
\centering
\includegraphics[trim=47 5 0 0mm,width=0.95\hsize,clip]{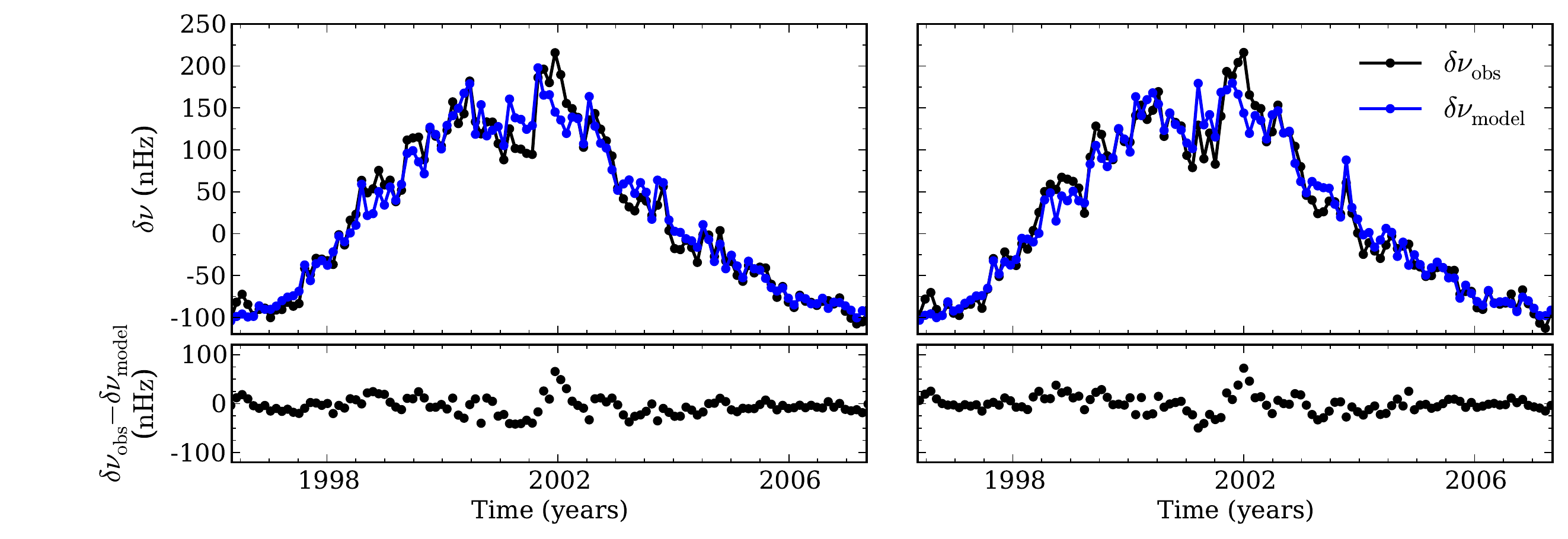}\vspace{-0.2cm}
\caption{Top: Comparison between the observed (black) and model (blue) frequency shifts. The model frequency shifts include a global long-term component related with the overall solar magnetic field and a spot-induced component which varies on short timescales (i.e. $\delta\nu_{\rm model}=\delta\nu_{\rm global}+\delta\nu_{\rm spots}$; Eq.~(\ref{eq:mfshifts})). Bottom: Residuals between the observed and model frequency shifts. Left and right panels concern each sample of independent data points.}
\label{fig:fshifts}
\end{figure*}

Figure~\ref{fig:fshifts} shows the comparison between the model predictions and the observational data for both samples of independent data points. Although the model and observed frequency shifts agree reasonably well and we have checked that our estimate of the relative contribution of each component is robust, the residuals shown in the bottom panels are found to vary in phase with the quasi-biennial variations in the frequency shifts \cite[e.g.][]{Fletcher2010,Broomhall2012}. This may suggest that our model is missing a mid-term contribution varying on a quasi-biennial timescale.

\section{Weighted sum of the frequency-shift differences}
\label{sec:WD}

The total area covered by sunspots varies on a timescale of days. Therefore, one may expect that sunspots induce frequency variations on the same timescale, being the main source for the 36-d frequency shifts. With this in mind, we propose a new quantity to investigate the short-term variations in the frequency shifts, being almost insensitive to the global long-term component of the frequency shifts \cite{Santos2016b}. This quantity combines the information contained in both observed frequency shifts and sunspot areas and is defined as the weighted sum of the frequency-shift differences ($W_{\rm D}$), i.e.
\begin{equation}
W_{\rm D}=\sum_n\Delta\delta\nu_n\times S_n,
\label{eq:WD}\end{equation}
where $\Delta\delta\nu_n$ is the frequency-shift difference at time $n$ ($\Delta\delta\nu_n=\delta\nu_n-\delta\nu_{n-1}$ with $n=2,3,4,...$) and $S_n$ is the weight to be determined by the variation in the total area covered by sunspots :
\begin{equation}
S_n= \begin{cases} 1, & \mbox{if }\Delta A_{{\rm T},\,n}>0 \\ -1, & \mbox{if }\Delta A_{{\rm T},\,n}<0 \end{cases}\end{equation}
where $\Delta A_{{\rm T},\,n}=A_{{\rm T},\,n}-A_{{\rm T},\,n-1}$ and $A_{{\rm T},\,n}$ is the total area covered by sunspots at a given time $n$. If the short-term variations in the frequency shifts and in the spot areas are completely correlated, $\Delta\delta\nu_n$ and $\Delta A_{{\rm T},\,n}$ will always have the same sign and $W_{\rm D}$ will correspond to the sum of the absolute (moduli) frequency-shift differences ($M_{\rm D}$)
\begin{equation}
M_{\rm D}=\sum_n|\Delta\delta\nu_n|.
\label{eq:MD}\end{equation}

\begin{figure*}[t]
\centering
\includegraphics[trim=59 43 0 0mm,width=0.95\hsize,clip]{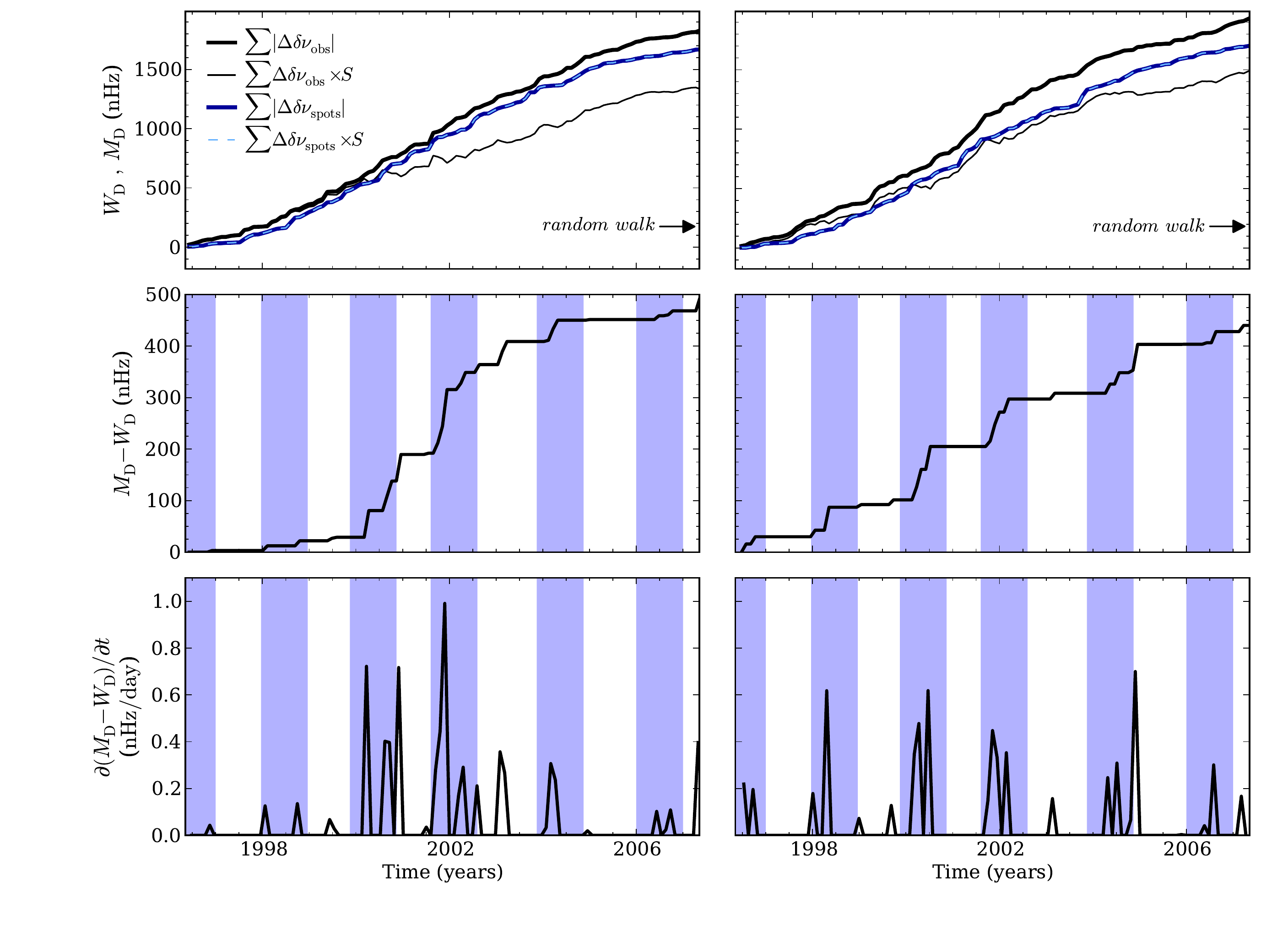}\vspace{-0.2cm}
\caption{Top: Quantities $M_{\rm D}$ (thick solid lines) and $W_{\rm D}$ (thin solid and dashed lines) for the observed (black) and spot-induced (dark and light blue) frequency shifts. The arrow marks the standard deviation for a random walk. Middle: Difference between the quantities $M_{\rm D}$ and $W_{\rm D}$. Bottom: Time derivative of the difference between $M_{\rm D}$ and $W_{\rm D}$. The blue bars are centered around the maxima of the quasi-biennial signal, having a width of 1 yr. Left and right panels concern each sample of independent data points.}
\label{fig:shortv}
\end{figure*}

The top panels of figure~\ref{fig:shortv} compare the weighted sum of the frequency-shift differences, $W_{\rm D}$, and the sum of the absolute values of the frequency-shift differences, represented by the quantity $M_{\rm D}$. For the spot-induced frequency shifts obtained in Sect.~\ref{sec:sfshifts}, $M_{\rm D}$ and $W_{\rm D}$ overlap. This shows that the spot-induced frequency shifts, combined as done for the observed frequency shifts, are mostly proportional to the sunspot area, being weakly dependent on the sunspot latitudes. For the observed frequency shifts, $M_{\rm D}$ and $W_{\rm D}$ do not overlap and the maximum difference between them corresponds to the amount of correlation that is lost over the cycle, which cannot be fully explained by the sunspots on the far-side of the Sun \cite{Santos2016b}. Nevertheless, the comparison of $W_{\rm D}$ with the expected standard deviation for a random walk (at the end of the cycle), shows that the short-term variations in the frequency shifts and in the total area covered by sunspots are strongly correlated.

The middle panels of figure~\ref{fig:shortv} show the difference between the quantities $M_{\rm D}$ and $W_{\rm D}$. To better identify the times when frequency shifts and spot areas behave differently, the time derivative of that difference, $\partial\left(M_{\rm D}-W_{\rm D}\right)/\partial t$, is shown in the bottom panels of figure~\ref{fig:shortv}. The times of loss of correlation correspond to $\partial\left(M_{\rm D}-W_{\rm D}\right)/\partial t\neq0$. We found that most of the loss of correlation between the short-term variations in the frequency shifts and in the sunspot areas occurs around epochs of maximum of quasi-biennial signal found by \cite{Broomhall2012}.

\section{Conclusions}
\label{sec:con}
We investigated the short-term variations in the solar acoustic frequencies, which we expect to be mainly induced by the presence of sunspots on the solar surface. 

We derived a parametric model for the spot-induced frequency shifts. The total model frequency shifts also account for a long-term component varying on the timescale of the solar cycle (11 yr). This component is assumed to be representative of the weak component of the magnetic field. The comparison between the model and observed frequency shifts indicates that the spot-induced contribution is $\sim30\%$ of the total frequency shifts. The remaining $\sim70\%$ corresponds to the long-term component. However, our results also suggest that there is a mid-term component varying on a quasi-biennial timescale that is not account for.

We also propose a new method for the analysis of the correlation between the frequency shifts and other activity proxies, such as the area coverage by spots. The method is based on the weighted sum of the frequency-shift differences, being the weight determined by the variation in the area covered by sunspots. This quantity amplifies the signal from the short-term component of the frequency shifts and is almost insensitive to the long-term component. We found a strong correlation between the short-term variations in the frequency shifts and in the sunspot areas. However, there is a significant loss of correlation at specific moments during the cycle, which tends to coincide with times of maximum of the quasi-biennial signal. These results further suggest that there is a mid-term contribution which has not been taking into account.\\

\begin{acknowledgement}
\hspace{-0.25cm} 
This work was supported by Funda\c{c}\~{a}o para a Ci\^{e}ncia e a Tecnologia (FCT) through the research grant UID/FIS/04434/2013. ARGS acknowledges the support from FCT through the Fellowship SFRH/BD/88032/2012 and from the University of Birmingham. MSC and PPA acknowledge support from FCT through the Investigador FCT Contracts No. IF/00894/2012 and IF/00863/2012 and POPH/FSE (EC) by FEDER funding through the programme Programa Operacional de Factores de Competitividade (COMPETE). TLC and WJC acknowledge the support of the UK Science and Technology Facilities Council (STFC). The research leading to these results has received funding from EC, under FP7, through the grant agreement FP7-SPACE-2012-312844 and PIRSES-GA-2010-269194. ARGS, MSC, and PPA are grateful for the support from the High Altitude Observatory (NCAR/UCAR), where part of the current work was developed.
\end{acknowledgement}

%
\bibliographystyle{woc}
\bibliography{shortv_proc}\vspace{-3.6cm}
\end{document}